\documentclass[DIV12]{scrartcl}

\usepackage{hyperref}
\usepackage[utf8]{inputenc}
\usepackage{tabularx}
\usepackage{setspace}
\usepackage{graphicx}
\usepackage{booktabs}
\usepackage{amsmath}

\hypersetup{
	colorlinks=true,
	linkcolor=black,
	citecolor=black,
	urlcolor=black
}

\date{}
\usepackage{authblk}

\usepackage{natbib}
\title{Audio spectrogram representations for processing with Convolutional Neural Networks}

\usepackage{fancyhdr}
\begin{document}

\author[1]{Lonce Wyse\thanks{lonce.wyse@nus.edu.sg}}

\affil[1]{\small National University of Singapore}

\maketitle

\thispagestyle{fancy}

\begin{abstract}
One of the decisions that arise when designing a neural network for any application is how the data should be represented in order to be presented to, and possibly generated by, a neural network.  For audio, the choice is less obvious than it seems to be for visual images, and a variety of representations have been used for different applications including the raw digitized sample stream, hand-crafted features, machine discovered features, MFCCs and variants that include deltas, and a variety of spectral representations.  This paper reviews some of these representations and issues that arise, focusing particularly on spectrograms for generating audio using neural networks for style transfer. 
\bigskip


\end{abstract}

\section{Introduction}

Audio can be represented in many ways, and which one is “best” depends on the application as well as the processing machinery. For many years, feature design and selection was a key component of many audio analysis tasks and the list includes spectral centroid and higher-order statistics of spectral shape, zero crossing statistics, harmonicity, fundamental frequency, and temporal envelope descriptions. Today, the general wisdom is to let the network determine the features it needs to accomplish its task. 

For classification, particularly in speech, Mel Frequency Cepstral Coefficients (MFCCs) which describe the shape of a spectrum, have a long history. Although they are a lossy representation, they are used for their classification and identification effectiveness even at very reduced data rates compared to sampled audio. MFCC's have also been used for environmental sound classification with convolutional neural networks ~\citep{Piczak:2015}, although the reported 65\% classification accuracy might be helped with a less lossy representation. Raw audio samples have also been used for event classification, for example in SoundNet ~\citep{aytar2016soundnet}.  

\section{Sound Representation for Generative Networks}
For generative applications, a representation that can be used to synthesize high-quality sound is essential. This rules out “lossy” representations such as MFCCs and many hand-crafted feature sets, but still leaves several options. 

Raw audio samples are lossless and trivially convertible to audio. WaveNet ~\citep{wavenet:2016}, is a deep convolutional net (not recurrent) that uses raw audio samples as input and is trained to predict the most likely next sample in a sequence. During the generative phase, each predicted sample is incorporated into the sequence used to predict the following sample. With “conditioning” information (such as which phoneme is being spoken) provided along with input, interesting parametric control at synthesis time is possible.  WaveNet implementations  run as deep as 60 layers, and raw audio is typically sampled at rates ranging from 16K to 48K per second, so synthesis is slow at many minutes of processing per second of audio.  

Magnitude spectra can also be used for generative applications given techniques for deriving phase from properties of the magnitude spectra to reconstruct an audio signal. The most often-used phase reconstruction technique comes from ~\citet{G&L}, which is implemented in the Librosa library ~\citep{mcfee2015librosa}. However, it involves many iterations of forward and inverse Short-time Fourier Transforms (STFTs), and is fundamentally not real time (the whole temporal extent of the signal is used to reconstruct each point in time), and is plagued by local minima in the error surface that sometimes prevent high-quality reconstruction. Recent research has produced methods that are theoretically and in practice real time ~\citep{RTISI}~\citep{pruvsa2016real}; methods that can produce very convincing transients (temporally compact events) ~\citep{pruvsatowards}; and non-iterative methods of reasonable quality that are as fast to compute as a single STFT ~\citep{SPSI}.   

Spectrograms are 2D images representing sequences of spectra with time along one axis, frequency along the other, and brightness or color representing the strength of a frequency component at each time frame. This representation is thus at least suggestive that some of the convolutional neural network architectures for images could be applied directly to sound.  

Style transfer ~\citep{GatysStyle} is a generative application that uses pre-trained networks to create new images combining the content of one image and the style of another.  Because of the plethora of image networks available (e.g. VGG-19 ~\citep{DBLP:journals/corr/SimonyanZ14a} pre-trained  on the 1.2M image database ImageNet ~\citep{Deng09imagenet:a}) and the dearth of networks trained on audio data,  the question naturally arises as to whether the image nets would be useful for audio style transfer representing audio spectrogram images. We ran some experiments with the pre-trained VGG-19 network, with the goal of superimposing “style” or textural features from one spectrogram on the “content” or structural features of another. The features were defined as in ~\citep{GatysStyle}, so that content features were just the activations in deeper layers of the network, and style features were defined as the Gram matrix, a second-order measure derived from activations on several shallower layers. 

In order to use spectral data for this purpose, several issues had to be addressed. Because image processing networks work on 3-channel RGB input, the single-channel magnitude values of the spectrograms must be duplicated across 3 channels to work with the pre-trained network. Since color channels are processed differently from each other in the neural network, the post-processing synthesized color image must be converted back to a single channel based on luminosity to be meaningful as a spectrogram.  

Although processing sonograms as images “works” in the sense that visual characteristics are combined in interesting nonlinear ways, the resulting sounds are not nearly as compelling as style transfer for visual images is. The issue is likely due to the difference between how sonic objects are represented in spectrograms compared to how visual objects are represented in 2D, and the way convolutional networks are designed to work with these images. 

Convolutional neural networks designed for images use 2D convolution kernels that share weights across both the x and the y dimensions. This is based in part on the notion of translational invariance, which means that an image feature or object is the same no matter where it is in the image. For sonic objects in the linear-frequency sonogram, this is true when objects are shifted in the x dimension (time), but not when they are shifted in the y dimension (frequency). Audio objects consist of energy across the frequency dimension, and as a sound is raised in pitch, its representation not only shifts up, but changes in spatial extent. A log frequency representation may go some way to addressing this issue, but the non-local distribution of energy across frequency of an audio object might still be problematic for 2D convolution kernels. Sound images also present other challenges compared to visual images - for example, sound objects are “transparent” so that multiple objects can have energy at the same frequency, where a given pixel in a visual image almost always corresponds to only one object. In addition, audio objects are non-locally distributed over a spectrogram whereas visual objects tend to be comprised of neighboring pixels in an image. 

Dmitry Ulyanov ~\citet{Ulyanov:2016} reports in a blog posting about using convolutional neural networks in a different way for audio style transfer. He uses spectrograms, but instead of representing the frequency bins as the y dimension in an image, he considers the different frequencies as existing at the same point in a 1D representation as stack of “channels” in the same way the 3 channels for red, green, and blue are stacked at each point in a 2D visual image. As in image applications, the convolution kernel spans the entire channel dimension;  there is no small shared-weight convolution kernel that shifts along the channel dimension as it does in the spatial dimensions. The number of audio channels, typically 256 or 512, is much greater than the 3 channels used for color images, and the vertical dimension is reduced to one. 

There are two remarkable aspects to the network used by Ulyanov for style transfer that differentiate it from the “classical” approach described by Gatys et al. ~\citep{GatysStyle}. First, the network uses only a single layer. The network activations driving content generation and those driving style generation come from one and the same set of weights. The difference between content and style thus comes not from the depth of the layers, but only from the difference between first-order and second-order measures of activation. Secondly, the network was not pre-trained, but uses random weights. The blog post claims this unintuitive approach generated results as good as any other, and the sound examples posted are indeed compelling. 

To further investigate the utility of spectrogram representations and the hypothesis that weights are unimportant for style transfer, a network with two convolutional layers and two fully-connected layers was trained on the ESC-50 data set ~\citep{Piczak:2015} consisting of 2000 5-second sounds. Sounds were represented as spectrograms consisting of 856 frames with 257 frequency bins, and the network was trained to recognize 50 classes. We then compared pre-trained and random weight values for style transfer\footnote{\label{network}The network was trained with 2 convolutional layers of 2048 and 64 channels resp., used relu activation functions, and each was followed by max pooling of size 2 with strides of 2. A fully connected final layer had 32 channels. A secondary classification was performed simultaneously (multi-task learning) as regularization, where sounds were divided into 16 balanced classes based on spectral centroid. Details and sound examples at http://lonce.org/research/audioST}. 

\begin{figure}[h]
\centering
\includegraphics[width=.85\textwidth]{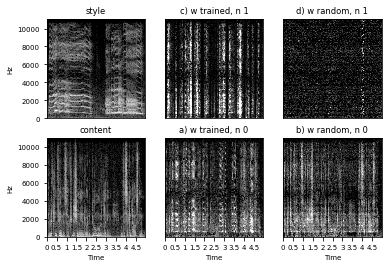}
\caption{a) With trained network weights and no added image noise, the result shows well-integrated features from both style and content. b) With random weights, style influence is hard to detect and content sounds noisy. c) Adding noise to the initial image results in sound that has the gross amplitude features of the content and a noisy timber barely identifiable with the style source d) Random weights and added image noise cause the loss of any sense of either content and style. }
\label{fig:st_net}
\end{figure}

Sonograms generated with different weight and noise conditions are shown in Figure \ref{fig:st_net}. The content target is speech and the style target is a crowing rooster. This study shows a significant difference between random and pre-trained weights. Additionally, the network trained for audio classification does not introduce the audible artifacts of the kind we found using an image-trained network. Although style transfer does work without regard to weights based only on the first-order and second-order content and style matching strategy, a network trained for audio classification appears to generate a more integrated synthesis of content and style.  

For the architecture we used, style suffers more than content from noise effects, whether added to the initial image, or in the form of random weights. Also, to compensate for the reduction of parameters in the network when arranging frequency bins as channels, it is necessary to dramatically increase the number of channels in the network layer(s) in order for longer timescale style features to appear in the synthesis. Ulyanov used 4096 channels, we used 2048 in the first layer. This is both greater than the typical channel depth used in image processing networks, and greater than was necessary to pre-train the network on the classification task.

\section{Summary}
Spectral representations may have a role in applications that use neural networks for classification or regression. They retain more information than most hand-crafted features traditionally used for audio analysis, and are of lower dimension than raw audio. The are particularly useful for generative applications due to available techniques for reconstructing high-quality audio signals. Linear-frequency sonograms can not be treated in the same was as images are by 2D convolutional networks, but other approaches such as considering frequency bins as channels are being explored and show promising results.

\bibliography{paper}

\end{document}